# Monte Carlo Study of Apparent Mobility Reduction in Nano-MOSFETs

K. Huet, J. Saint-Martin, A. Bournel,
S. Galdin-Retailleau and P. Dollfus

Institut d'Electronique Fondamentale
IEF, Univ. Paris Sud, CNRS
Orsay, France
karim.huet@ief.u-psud.fr

G. Ghibaudo and M. Mouis

Institut de Microélectronique, Electromagnétisme et Photonique
IMEP, CNRS-INPG-UJF
Grenoble, France

*Abstract* — The concept of mobility is discussed in the case of unstrained and strained nanoscale DG MOSFET thanks to particle Monte Carlo device simulation. Without the introduction of specific scattering phenomenon for short channel devices, the apparent mobility extracted from simulated electrical characteristics decreases with the shrinking of the channel length, as experimentally observed elsewhere. We show that this reduction at room temperature is caused by non stationary effects. Moreover, both simulation results and experimental data may be well reproduced by a Mathiessen-like model, using a "ballistic mobility" extracted from MC simulations together with the usual long channel mobility.

## I. INTRODUCTION

The concept of mobility, resulting from an analysis of stationary transport where carrier velocity is limited by scattering phenomena, has been widely used till today in microelectronics as a measurable factor of merit and as a parameter of analytical models developed to predict device performance. If scatterings are still playing a major role in decananometer MOSFET and cannot be neglected [1-3], ballistic transport in the channel takes a growing importance as the gate length of MOSFETs tends to the nanometer scale. In this context, the mobility concept may appear as highly questionable.

Dramatic reduction of the mobility measured at short gate length has been observed by different techniques [4-9]. The understanding of such a behavior is raising a growing and challenging debate. Doping pockets, charge and neutral defects are suspected of limiting mobility when MOSFET are scaled down. However, the assumption that the measured mobility in nanoscale devices is actually the physical mobility of the channel can be questioned. The observed mobility decrease could be only apparent and mainly the consequence of non stationary effects. In this context, the use of analytical and compact models can provide useful clues [10-12] for this debate. Using a model proposed by Shur to bridge the gap between the usual analytical I-V models for FET transistors, based on the notion of stationary mobility, and the physics of quasi ballistic transport in nanoscale devices [10], Lusakowski *et al.* observed that the mobility extracted from experimental I-V characteristics of nanometer-scaled bulk MOSFET may be limited by ballistic effects [5]. According to Shur's approach in [11], the apparent or effective mobility $\mu_{eff}$ may be given by the "Mathiessen-like" following relation:

$$\mu_{eff} = \frac{1}{\mu_{long}^{-1} + \mu_{bal}^{-1}}, \quad (1)$$

where $\mu_{long}$ is the long channel mobility and $\mu_{bal}$ is the "ballistic mobility" as defined by Shur. As stressed in [11], the $\mu_{bal}$ parameter is just an apparent mobility, since the mobility concept is related to the collision-dominated regime.

Using our semi-classical Monte Carlo (MC) device simulator MONACO, we explore the concept of mobility in N-channel double-gate (DG) MOSFETs. The dominant influence of non stationary transport in these devices will be highlighted and its effect on apparent mobility will be investigated using an approach similar to that proposed by Shur. The resulting model will be confronted to MC simulations and experiment.

## II. STUDIED DEVICES AND EXTRACTION METHOD

MONACO is a particle MC device simulator coupled with 2D or 3D Poisson solver. In this work, we consider scattering mechanisms related to phonons, ionized impurities and $SiO_2$/Si surface roughness. The latter is accounted for via an empirical combination of diffusive and specular reflections (14% of diffusive reflections). We use an analytical description of the conduction band structure with 6 ellipsoidal Δ valleys (see for example [2] and references therein for more details on MONACO). Quantization effects have not been taken into account here. However, results obtained with a multi subband simulator do not show significant differences in ballistic effects and I-V characteristics for body thickness $T_{Si}$ down to 5 nm [13]. The channel length $L_{ch}$ of the studied DG structures ranges from 10 nm to 400 nm. The body thickness $T_{Si}$ is equal to 5 nm and the gate oxide thickness $T_{ox}$ (or equivalent oxide thickness, EOT) is 1.2 nm.

This work was supported by the French Agence Nationale de la Recherche through project MODERN (ANR-05-NANO-002) and by the European Community 6th FP under contracts IST-026828 (IP PULLNANO) and IST-506844 (NoE SINANO).

Two series of devices have been simulated: DG MOSFETs in which standard scattering rates are considered in the channel and artificial structures with the same geometry but with a ballistic channel, i.e. without any scattering events. The channel is non intentionally doped. The temperature is T = 300 K, unless otherwise stated. We consider a metallic midgap gate material. Some strained devices have also been investigated.

A mobility $\mu_{extr}$ is extracted from drain current $I_D$ variations as a function of the gate voltage $V_{GS}$ at low drain voltage $V_{DS}$ using a method inspired by Hamer [14], in accordance with experimental characterization techniques. The following formula (2) is used to fit the MC results:

$$I_D = C_{ox} \frac{W}{L_{ch}}(V_{GS} - V_T)\frac{\mu_{extr}}{1+\theta(V_{GS} - V_T)}V_{DS}. \quad (2)$$

The gate capacitance $C_{ox}$ is given by the analysis of MC results [15]. The threshold voltage $V_T$, the extracted mobility $\mu_{extr}$ and the gate field reduction parameter $\theta$ are used as fitting parameters.

## III. RESULTS AND DISCUSSION ON APPARENT MOBILITY REDUCTION

Fig. 1 shows the $I_D(V_{GS})$ curves obtained at low $V_{DS}$ (0.05 V) for unstrained DG MOSFETs when the scattering phenomena are taken into account. For the sake of comparison, results concerning devices with ballistic channel are reported in the inset of fig. 1. The shrinking of the channel length does not affect the electrical characteristics, except for the 10 nm-long devices where short channel effects (SCE) become significant because of a too small $L_{ch}/T_{Si}$ ratio [15]. When scattering is included in the channel, the reduction of $L_{ch}$ induces a continuous increase of $I_D$ at a given bias point. The extracted $V_T$ is found almost constant (within the 0.289 V – 0.295 V range), except for the 10 nm device which suffers from SCE ($V_T = 0.269$ V). $V_T$ variations are therefore clearly not responsible for the $I_D$ enhancement at short $L_{ch}$. The $V_T$-$L_{ch}$ behavior is similar for devices with a ballistic channel. The extracted mobilities for diffusive ($\mu_{extr\_scat}$) and ballistic ($\mu_{extr\_bal}$) channel devices are shown on fig. 2. The decreasing behavior of $\mu_{extr\_scat}$ with $L_{ch}$ is similar to experimental observations [4-9]. As the interaction mechanisms considered are identical for all $L_{ch}$, this apparent mobility reduction is not necessarily due to additional scattering phenomena likely to have an increasing influence in short channel devices, as it ha been suggested.

In order to quantify the role of stationary effects, we use the well known parallel field corrected mobility [16]:

$$\mu_{long}^* = \frac{1}{\frac{1}{\mu_{long}} + \frac{V_{DS}}{v_{sat}.L_{ch}}}, \quad (3)$$

where $v_{sat}$ is the saturation velocity of Si. The $v_{sat}$ and $\mu_{long}$ values extracted from MC simulations are reported in table 1. As shown in fig.2, while the parallel field corrected mobility (dotted line) is reduced at short channel length, this correction alone cannot reproduce quantitatively $\mu_{extr\_scat}(L_{ch})$.

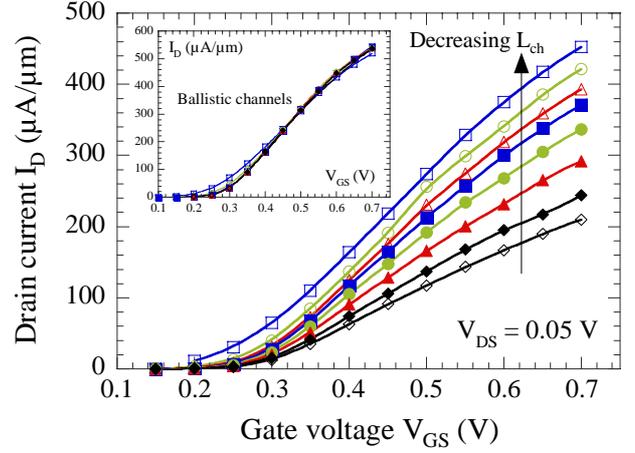

Figure 1. $I_D(V_{GS})$ curves at low $V_{DS}$ for different $L_{ch}$ (10, 15, 20, 25, 35, 50, 75 and 100 nm) of unstrained devices. Inset: devices with a ballistic channel.

To highlight the strongly out-of-equilibrium transport in the studied devices, we analyzed the population of electrons flowing through the channel from source-end to drain-end. When decreasing $L_{ch}$, the fraction of ballistic electrons, that is called the "intrinsic ballisticity" $B_{int}$, plotted in the inset of fig. 2, increases. This effect contributes to $I_D$-enhancement and makes the drain current closer and closer to the ballistic limit. Thus, the "effective ballisticity" $B_{eff}$, which is the ratio $I_{Dscat}/I_{Dbal}$ where $I_{Dscat}$ (resp. $I_{Dbal}$) is the current obtained in a channel with (resp. without) scattering, increases also as $L_{ch}$ decreases. These results are similar to those obtained in a previous work where on-state device operation ($V_{GS} = V_{DS} = 0.7$ V) was studied [2].

Therefore, since no $L_{ch}$-dependent scattering mechanism is considered in our simulations, since parallel field effects do not reproduce correctly the extracted mobility degradation and since ballistic transport becomes significant in short scaled devices, non stationary effects must be taken into account in mobility models.

## IV. MOBILITY REDUCTION AND "BALLISTIC MOBILITY"

The model developed by Shur introduces a parameter $\mu_{bal}$, homogeneous to a mobility and proportional to channel length, rewritten here as:

$$\mu_{bal} = K_{bal}.L_{ch}. \quad (3)$$

TABLE I. MODEL PARAMETERS DEDUCED FROM MC SIMULATIONS

| Simulated devices | $\mu_{long}^a$ $(cm^2.V^{-1}.s^{-1})$ | $K_{bal}^b$ $(cm^2.V^{-1}.s^{-1}.nm^{-1})$ | $v_{sat}^c$ $(\times 10^7 cm.s^{-1})$ |
|---|---|---|---|
| Unstrained, T = 300 K | 510 | 8.4 | 1.0 |
| Unstrained, T = 400 K | 355 | 7.2 | 1.0 |
| Strained, T = 300 K | 1000 | 13.2 | 1.2 |

[a] from 5 nm-thick long-channel devices MC, [b] from ballistic channel devices MC [c] from bulk material in a constant field MC.

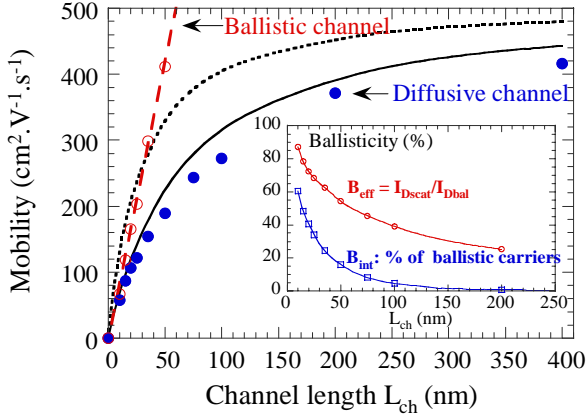
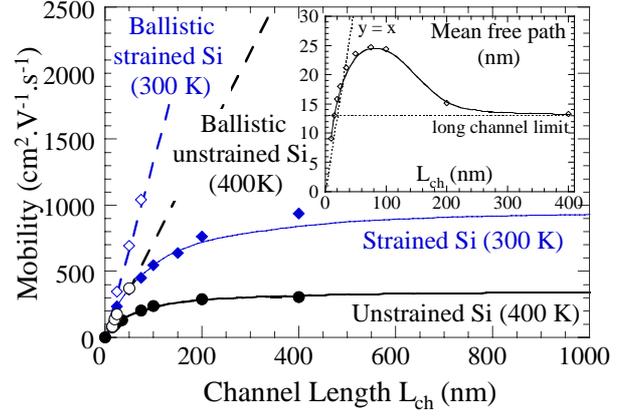

Figure 2. Mobility extracted from $I_D(V_{GS})$ curves at low $V_{DS}$ as a function of channel length $L_{ch}$ (symbols) for unstrained devices. Dotted line: Field corrected stationary mobility. Dashed line: Linear fit for ballistic channel devices. Continuous line: "Matthiessen-like" formalism. Inset: Intrinsic and effective ballisticies ($B_{int}$ and $B_{eff}$) vs. $L_{ch}$ at $V_{DS} = 0.05$ V and $V_{GS} = 0.7$ V.

Figure 3. Mobility extracted from $I_D(V_{GS})$ curves at low $V_{DS}$ as a function $L_{ch}$ (symbols). Continuous line: "Matthiessen-like" model. Dashed line: linear fits for ballistic channel devices. Insert: mean free path $\lambda_{mfp}$ extracted from MC simulations [17] as a function of $L_{ch}$.

This parameter $\mu_{bal}$ is equivalent to the apparent mobility $\mu_{extr\_bal}$, extracted from MC simulation of devices with a ballistic channel. Indeed, as shown on fig. 2 and 3, $\mu_{extr\_bal}$ is proportional to $L_{ch}$. The slopes $K_{bal}$ extracted for the different studied devices are reported in table 1. Moreover, asymptotic behavior of $\mu_{extr\_scat}(L_{ch})$ is consistent with (1): $\mu_{extr\_scat}$ is close to $\mu_{extr\_bal}$ at small $L_{ch}$, then increases slower than $\mu_{extr\_bal}$ as $L_{ch}$ increases, and finally saturates at higher $L_{ch}$. This shows that a Shur-based model can be useful to describe the apparent decrease of extracted mobility with $L_{ch}$. Shur defines $K_{bal}$ as:

$$K_{bal} = \frac{2.q}{\pi.m.v_{th}}, \quad v_{th} = \sqrt{\frac{8.k_B.T}{\pi.m}}, \quad (4)$$

where q is the electron charge, m the conduction effective mass, $v_{th}$ the average thermal velocity, $k_B$ the Boltzmann constant and T the temperature [9].

Using (4) for unstrained (resp. strained) devices at T = 300 K, the value of $K_{bal}$ is found to be equal to 20 (resp. 23.9) $cm^2.V^{-1}.s^{-1}.nm^{-1}$ with an average effective mass m = 0.26 $m_0$ (resp. 0.19 $m_0$) where $m_0$ is the free electron mass. Compared to the values deduced from MC simulation, the $K_{bal}$ value calculated from (4) is overestimated.

If the usual expression of $\mu_{long}$ is used [11]:

$$\mu_{long} = \frac{2.q.\lambda_{mfp}}{\pi.m.v_{th}}, \quad (5)$$

the formalism gives unrealistic values of the mean free path $\lambda_{mfp}$. Indeed, substituting (3), (4) and (5) into (1) we can write:

$$\mu_{eff} = \frac{\mu_{long}}{1 + \frac{\lambda_{mfp}}{L_{ch}}}, \quad (6)$$

the long channel mobility and the mean free path can be deduced by adjusting (6) to the extracted mobility. For unstrained devices at T = 300 K, a satisfactory fit (not shown) was obtained with $\mu_{long} = 502$ $cm^2.V^{-1}.s^{-1}$ and $\lambda_{mfp} = 79$ nm.

The obtained mean free path is unexpectedly high compared to those extracted from MC simulation [17], plotted in the inset of fig. 3. Besides, calculating $\mu_{long}$ using (5) with this value of $\lambda_{mfp}$ gives a value of 1610 $cm^2.V^{-1}.s^{-1}$, which is 3 times higher than both the fitted value and the one issued from long channel MC simulation (reported in table 1).

Furthermore, on the basis of experiments, Andrieu [9] highlighted a similar inconsistency of the model. Lusakowski et al. [5] used a new definition of $\lambda_{mfp}$ involving doping pockets to find a consistent value of $\mu_{long}$. However, the devices studied in the present work having an undoped channel with well defined source and drain junctions, the possible influence of pockets is not relevant.

Seeing that a new degree of freedom is needed to have a consistent model, another approach retaining the main features of the Shur model is investigated. Combining (3) with the initial "Mathiessen-like" relationship (1), we can rewrite the effective mobility as:

$$\mu_{eff} = \frac{\mu_{long}}{1 + \frac{\mu_{long}}{K_{bal}.L_{ch}}}. \quad (7)$$

In this model, $\mu_{long}$ is an easily accessible value through long device simulations (or measurements) and $K_{bal}$ can be deduced from simulation of ballistic channel devices. Using the calculated $K_{bal}$ and the long channel MC mobility $\mu_{long}$ reported in table 1, the resulting plot of (7) on fig. 2 and 3 (continuous lines) shows the good agreement of the model with MC simulations of the DG MOSFETs for all the studied conditions.

To further validate this approach, we apply (7) on experimental data for 14 nm-thick unstrained FD-SOI MOSFETs [9]. In fig. 4, we used (7) with $\mu_{long} = 430$ $cm^2.V^{-1}.s^{-1}$ (i.e. mobility of the 5 $\mu$m-long channel device in [9]) and $K_{bal} = 8.4$ $cm^2.V^{-1}.s^{-1}.nm^{-1}$ (predicted by MC simulations of DG MOSFETs with a ballistic channel, see table 1).

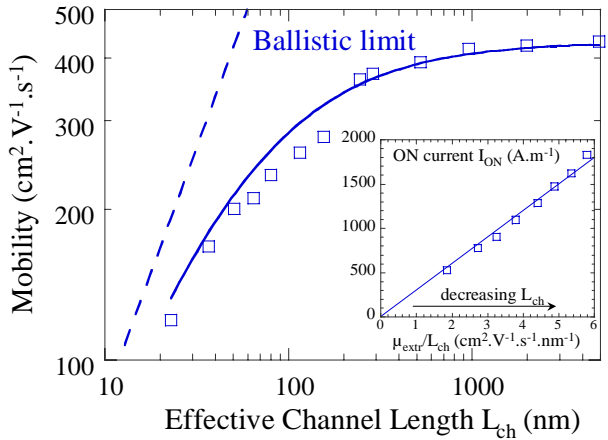

Figure 4. Experimental mobility (symbols) [9]. Continuous line: "Mathiessen-like" formalism. Dashed line: ballistic channel limit ($K_{bal}$ = 8.4 cm$^2$.V$^{-1}$.s$^{-1}$). Inset: ON state current $I_{ON}$ vs. $\mu_{extr}/L_{ch}$ ratio.

Knowing that no parallel field correction is used, this close agreement between the model and experimental measurements means that the mobility degradation observed when $L_{ch}$ is reduced can be mostly explained by non stationary transport.

Furthermore, it means that, in the studied cases, the "ballistic mobility" does not depend on the device geometry. It suggests that $K_{bal}$ extracted from MC simulation can be quantitatively used to interpret experimental data within a Shur-like model as long as SCE remain negligible.

The apparent mobility extracted from $I_D(V_{GS})$ curves is not a real mobility according to the common definition of the term. However, this apparent mobility can still be used as an accurate factor of merit for the ON state of the transistor. Indeed, as shown in the inset of fig. 4, the ON current, obtained from MC simulated devices at $V_{DS} = V_{GS} = 0.7$ V, and the $\mu_{extr}/L_{ch}$ ratio, extracted from MC simulations at $V_{DS} = 0.05$ V, are linearly correlated.

## V. CONCLUSION

At room temperature, the often reported degradation of the apparent mobility extracted from experimental I-V characteristics as the gate length decreases may be explained mostly by non stationary effects. The occurrence of additional detrimental scattering phenomena (doping pockets, charge and neutral defects) in short channels is not obvious.

Besides, a model based on only 2 parameters, $\mu_{long}$ and $K_{bal}$ (only qualitatively provided by Shur's analytical expression), deduced from independent MC simulations, was shown to be able to reproduce correctly the extracted mobility behavior as a function of channel length for the simulated devices and for experimental measurements.

Therefore, the notion of apparent "ballistic mobility" and its use in a "Mathiessen-like" rule is relevant for the studied transistors.

ACKNOWLEDGMENT

The authors would like to thank D. Querlioz, A. Cros, W. Chaisantikulwat, R. Clerc, F. Andrieu and T. Poiroux for fruitful discussions and valuable suggestions.